\documentclass[reprint,superscriptaddress,amsmath,amssymb,aps,prd,nofootinbib,showpacs]{revtex4-1}

\usepackage{graphicx}
\usepackage{epstopdf}
\usepackage{textcomp}

\usepackage{dsfont}
\usepackage{amsfonts}
\usepackage{amsmath}
\usepackage{amssymb}
\usepackage[colorlinks=false]{hyperref}

\def\bkR{{\rm I\kern-.17em R}}
\def\bkC{{\rm \kern.24em \vrule width.05em height1.4ex depth-.05ex \kern-.26em C}}

\def\ev{\mbox{eV}}

\def\tev{\mbox{TeV}}

\def\be{\beta}

\def\vev#1{\left\langle {#1}\right\rangle}

\def\lsim{\mathrel{\rlap{\lower4pt\hbox{\hskip1pt$\sim$}}
    \raise1pt\hbox{$<$}}}
\def\gsim{\mathrel{\rlap{\lower4pt\hbox{\hskip1pt$\sim$}}
    \raise1pt\hbox{$>$}}}
\def\sqr#1#2{{\vcenter{\vbox{\hrule height.#2pt
         \hbox{\vrule width.#2pt height#1pt \kern#1pt
         \vrule width.#2pt}
         \hrule height.#2pt}}}}

\def\laq{\raise 0.4 ex \hbox{$<$}\kern -0.8 em\lower 0.62 ex\hbox{$\sim$}}
\def\gaq{\raise 0.4 ex \hbox{$>$}\kern -0.7 em\lower 0.62 ex\hbox{$\sim$}}

\def\be{\begin{equation}}
\def\ee{\end{equation}}
\def\beqa{\begin{eqnarray}}
\def\eeqa{\end{eqnarray}}

\def\dalemb#1#2{{\vbox{\hrule height.#2pt
        \hbox{\vrule width.#2pt height#1pt \kern#1pt \vrule width.#2pt}
        \hrule height.#2pt}}}

\def\dalemb#1#2{{\vbox{\hrule height.#2pt
        \hbox{\vrule width.#2pt height#1pt \kern#1pt \vrule width.#2pt}
        \hrule height.#2pt}}}

\def\gtorder{\mathrel{\raise.3ex\hbox{$>$}\mkern-14mu
             \lower0.6ex\hbox{$\sim$}}}
\def\ltorder{\mathrel{\raise.3ex\hbox{$<$}\mkern-14mu
             \lower0.6ex\hbox{$\sim$}}}

\newcommand{\dP}[2]{\frac{\partial #1}{\partial #2}}
\newcommand{\dT}[2]{\frac{d #1}{d #2}}

\begin{document}


\title{Two-scalar-field model for the interaction of dark energy and dark matter}

\author{Orfeu Bertolami}
\email{orfeu.bertolami@fc.up.pt}
\homepage{http://web.ist.utl.pt/orfeu.bertolami/}
\altaffiliation[Also at ]{Instituto de Plasmas e Fus\~ao Nuclear}
\affiliation{Departamento de F\'isica e Astronomia, Faculdade de Ci\^encias da Universidade do Porto \\
Rua do Campo Alegre 687, 4169-007 Porto, Portugal}

\author{Pedro Carrilho}
\email{pedro.carrilho@ist.utl.pt}
\affiliation{Instituto de Plasmas e Fus\~ao Nuclear, Instituto Superior T\'ecnico da Universidade T\'ecnica de Lisboa\\
Avenida Rovisco Pais 1, 1049-001 Lisboa, Portugal}

\author{Jorge P\'aramos}
\email{paramos@ist.edu}
\homepage{http://web.ist.utl.pt/jorge.paramos/}
\affiliation{Instituto de Plasmas e Fus\~ao Nuclear, Instituto Superior T\'ecnico da Universidade T\'ecnica de Lisboa\\
Avenida Rovisco Pais 1, 1049-001 Lisboa, Portugal}

\date{\today}

\begin{abstract}{
In this paper, we study the effects of an interaction between dark matter and dark energy through a two scalar field model with a potential $V(\phi,\chi)=e^{-\lambda\phi}P(\phi,\chi)$, where $P(\phi,\chi)$ is a polynomial. We show that the cosmic expansion dynamics of the Universe is reproduced for a large range of the bare mass of the dark matter field and that there exist solutions with transient accelerated expansion. A modification in the exponential behavior of the potential is studied, with important physical implications, including the possibility of more realistic transient acceleration solutions.
}
\end{abstract}


\maketitle

\section{Introduction}

A large number of models has been proposed to explain the dark sector of the Universe (see Refs. \cite{DE_rev,DM_rev} for reviews on dark energy (DE) and dark matter (DM), respectively). Most of those models assume that the dark components are noninteracting and treat them as fluids. However, there are neither theoretical arguments forbidding such an interaction, nor there exist sufficient observational results to rule it out. It is then just natural to study the general situation in which dark matter and dark energy are coupled, in order to gain a deeper insight into the nature of these components\footnote{Self-interacting dark matter has been discussed in Refs. \cite{ben1,ben2}.}. This is motivated by the theoretically appealing idea that the full dark sector can be treated in a single framework. Moreover, the fact that the present energy densities of dark energy and matter are observed to be the same order of magnitude, suggests a connection between them. 

An intimate connection between DE and DM is naturally expected in unification models, such as for instance the Chaplygin gas model and its generalizations \cite{KamMosPas,BilTupVio,BenBerSen}. Actually, in the context of this model, an assumption about the equation of state (EOS) of DE allows to extract an explicit interaction between the dark components \cite{BenBerSen2}. A map between the generalized Chaplygin gas (GCG) model and the interaction model to be discussed in the following paragraph can be found in Ref. \cite{OB_FGP_MLD}.

A general way to describe the DM-DE interaction is to introduce an energy exchange term $Q$ in the conservation equations as follows:
\begin{gather}
\dot\rho_{de}+3H(\rho_{de}+p_{de})=Q~,\ \dot\rho_{dm}+3H\rho_{dm}=-Q~.
\end{gather}
One may phenomenologically study this interaction by withholding any assumptions about the nature of the dark sector and treat it straightforwardly as a two-component fluid. The coupling $Q$ is usually taken to be of the form $Q=\delta_{de} H\rho_{de}+\delta_{dm} H\rho_{dm}$, where $H$ is the expansion rate and $\delta_i$ are coupling terms. This treatment is encountered in many observational studies \cite{OB_FGP_MLD,GuoOhTsu}.

An alternative path to study the interaction assumes that dark energy can be described by a scalar field $\phi$ in interaction with a fluid, the so-called interacting quintessence model: this is the case for the models of Refs. \cite{Wett,Amen}, in which the coupling is chosen to be $Q=f(\phi)\dot\phi\rho$, where $f(\phi)$ is a generic function. A similar mechanism is the so-called chameleon model \cite{KhoWel}. In these cases the field interacts with every component of the Universe, leading to observable effects in solar system tests of gravity. A simple modification is possible using a much smaller coupling for baryons than for DM as in Ref. \cite{DamGibGun}, or simply substituting $\rho$ with $\rho_{dm}$, so that DE couples only to DM. That is the case for the model in Ref. \cite{ZimPavChi}, in which the quintessence potential and the interaction term are derived from scaling assumptions.

A more fundamental approach to tackle the interaction treats DE and DM as fields, which abandons the need for fluids in the treatment of these components. Usually this is achieved through two new scalar fields, $\phi$ for DE and $\chi$ for DM \cite{FarPee,HueWan,MichAbdWang}. An interaction potential $V_{\text{int}}(\phi,\chi)$ is then introduced to account for the energy exchange. Ultimately these models also lead to the coupling of the interacting quintessence scenario, as long as the interaction is in the form of a DM mass term. This is to be expected, as will be derived in the following sections. These are then similar to the so-called VAMP models \cite{AndCar2} in which a particle is introduced whose mass varies with the quintessence field\footnote{These models can also involve fermions such as the neutrinos, in the so-called MaVaN models, see Ref. \cite{BernBert} and Refs. therein.}.

The advantages of this type of approach are manifold: the full set of coupled equations can be found from an action and consequently the functional form of the EOS parameter and the DE-DM coupling is fully determined. Furthermore, this is an elegant and straightforward way to link DM and DE with more fundamental physics models from which these components might stem.

A relevant property of these interaction models is the existence of an extra force between dark matter particles, not present in noninteracting models. This force can influence structure formation, creating an extra bias between baryon and dark matter fluctuations, which may in principle be measured through tests of the equivalence principle \cite{AmenToc,OB_FGP_MLD}. Furthermore, the varying dark matter mass can also have an influence in the anisotropy spectrum of the CMB since, among other effects, it may alter the ratio between dark matter and radiation densities at last scattering. This severely constrains the simplest interaction models, in which the DM mass is a linear function of the DE field, but its effects are still to be studied in detail for more complex models \cite{Hof}.

In this paper we present a scalar field interaction model with an interacting potential $V(\phi,\chi)$, that incorporates features of some quintessence models inspired in fundamental physics theories. We then characterize the physical solutions and ascertain the role of the interaction term in the cosmological evolution. The main equations of the model are derived in Sec. \ref{IntMod}. Their numerical solutions are presented in Sec \ref{Res}, along with the relevant physical results. Section \ref{Con} concludes with a discussion of the obtained results and a brief outlook on future developments.

\section{Interaction Model}\label{IntMod}

\subsection{Basic equations}

We consider two interacting canonical real scalar fields $\phi$ and $\chi$, whose Lagrangian density is given by\footnote{We use the $(-,+,+,+)$ metric signature and natural units with $\hbar=c=8\pi G=1$. As a consequence all masses come in terms of the reduced Planck mass, $M_p\equiv M_{Pl}/\sqrt{8\pi}$.}
\begin{gather}
\mathcal{L}_d=-\frac 12 g^{\mu\nu}(\partial_\mu\phi\partial_\nu\phi+\partial_\mu\chi\partial_\nu\chi)-V(\phi,\chi)~.
\end{gather}
Guided by the cosmological principle, we assume the geometry of the Universe to be given by a flat Robertson-Walker metric with line element
\begin{gather}
ds^2=-dt^2+a^2(t)\left(dr^2+r^2d\Omega^2_{S^2}\right)~,
\end{gather}
with $a(t)$ being the scale factor, normalized so that at present $a(t_0)=1$ for convenience, and $d\Omega^2_{S^2}$ the line element for the 2-dimensional sphere $S^2$. For the same reason, we consider both scalar fields to be homogeneous and isotropic, leading to the following field equations:
\begin{gather}
\ddot\phi+3H\dot\phi+\dP V\phi=0\label{phi}~,\\
\ddot\chi+3H\dot\chi+\dP V\chi=0\label{chi}~,
\end{gather}
where $H=\dot a/a$ is the expansion rate. From the stress-energy tensor we obtain the usual expressions for the pressure and energy density,
\begin{gather}
\rho_{d}=\frac 12\dot\phi^2+\frac 12\dot\chi^2+V(\phi,\chi)~,\\
p_{d}=\frac 12\dot\phi^2+\frac 12\dot\chi^2-V(\phi,\chi)\nonumber ~.
\end{gather}
We also consider a Universe filled with perfect fluids for matter and radiation, which we consider to be uncoupled and consequently to evolve as $\rho_m\propto a^{-3}$ and $\rho_r\propto a^{-4}$, respectively. Finally, we consider the Friedmann equation,
\begin{gather}
\label{fried}
H^2=\frac 13\left(\rho_m+\rho_r+\frac 12 \dot\phi^2+\frac 12 \dot\chi^2+V(\phi,\chi)\right)~.
\end{gather}
We introduce the density parameters $\Omega_i=\rho_i/3H^2$, in terms of which the usual deceleration parameter reads
\begin{gather}
q\equiv-\frac{\ddot aa}{\dot a^2}=\frac 12 \left(1+\Omega_r+3w_d\Omega_d\right)~,
\end{gather}
where $w_d=p_d/\rho_d$ is the EOS parameter for the fields.

\subsection{Interaction Potential}

We shall be interested in studying the following potential:
\begin{gather}
V(\phi,\chi)=e^{-\lambda\phi}P(\phi,\chi)+\frac 12 m^2\chi^2~,
\end{gather}
where $P(\phi,\chi)$ is a polynomial in $\phi$ and $\chi$ and $m$ is the dark matter bare mass. Such exponential couplings are inspired from fundamental theories like string or M-theory, or $N=2$ supergravity in higher dimensions \cite{Choi,SalSez,CopNunRos}. Notice that the interaction of chiral superfields in the context of N=1 supergravity inflationary models \cite{OvrSte,BerRos} has many common features with the present model. Furthermore, the exponential term for dark energy is the simplest way to vary its contributions from very high energy to the present and to respect the bond for nucleosynthesis \cite{BeaHanMel}. Hence under these conditions, one considers the general interaction term with an exponential multiplied by a polynomial of $\phi$ and $\chi$.

The polynomial $P(\phi,\chi)$ can be separated into the interacting and noninteracting terms, $P(\phi,\chi)=P_\phi(\phi)+P_{int}(\phi,\chi)$. For the noninteracting part, we choose the potential first studied in Ref. \cite{AlbSko},
\begin{gather}
P_\phi(\phi)=A+(\phi-\phi_0)^2~.
\end{gather}
As for the interacting term, an obvious choice is to require the $\chi$ field to be equivalent to a fluid of nonrelativistic matter, i.e. with negligible pressure. We recall that, according to Ref. \cite{Turner}, scalar field oscillations in a potential $V(\chi)=a\chi^n$ with frequency (i.e. mass) much greater than the expansion rate $H$, behave like a fluid with an average EOS given by
\begin{equation}
\label{eos}
\vev{p_\chi}=\frac{n-2}{n+2}\vev{\rho_\chi}~.
\end{equation}
Notice that this is equivalent to the virial theorem for power-law potentials, i.e. $\vev{\frac 12\dot\chi^2}=\frac n2\vev{ V(\chi)}$. Thus, in order to ensure that $\chi$ is pressureless at all times, we must set $n=2$. A main feature of the present model is that the interaction with the field $\phi$ leads to an oscillation with varying frequency. Note however that, unlike preheating models (see for example Ref. \cite{KofLinSta}), which exhibit parametric resonance, the frequency here changes slowly. In this case the computations discussed in Ref. \cite{Turner} for a constant frequency hold. To finish our discussion of the potential we rewrite it with an explicit DM varying mass term in terms of the DE field,
\begin{gather}
V(\phi,\chi)=V_{de}(\phi)+V_{dm}(\phi,\chi)~,
\end{gather}
with
\begin{gather}
\label{model}
V_{de}(\phi)=e^{-\lambda\phi}\left(A+(\phi-\phi_0)^2\right)~,\\
V_{dm}(\phi,\chi)=\frac 12 M^2(\phi)\chi^2\nonumber ~,
\end{gather}
where the mass function $M^2(\phi)$ is given in our model by $M^2(\phi)=m^2+2\tilde P(\phi)e^{-\lambda\phi}$ and the polynomial for $\tilde P$ is written as
\begin{gather}
\tilde P(\phi)=B+C\phi+D\phi^2~,
\end{gather}
where $B$, $C$ and $D$ are order unit parameters in terms of the appropriate powers of the reduced Planck mass. 

\subsection{Average Evolution Equations}

Given the high frequency of the oscillations, it is rather infeasible to integrate the $\chi$ equation numerically. For that reason we consider only averages of the field. In particular, we shall derive the equation for the dark matter density and work with that instead. First, we define the dark matter density and pressure from the EOS found in Eq. \eqref{eos}, 
\begin{gather}
\rho_{dm}=\frac 12 \dot\chi^2+\frac 12 M^2(\phi)\chi^2 ~,\\
p_{dm}=\frac 12 \dot\chi^2-\frac 12 M^2(\phi)\chi^2\nonumber ~.
\end{gather}
By construction, their averages over an oscillation cycle read
\begin{gather}
\vev{\rho_{dm}}=\vev{\dot\chi^2}=M^2(\phi)\vev{\chi^2}\label{avechi}~~,~~
\vev{p_{dm}}=0~.
\end{gather}
Next, we multiply Eq. \eqref{chi} by $\dot\chi$ and insert a term $\dot\phi V'_{dm}(\phi)$ to obtain
\begin{gather}
\dT{}{t}\left(\frac 12\dot\chi^2+V_{dm}(\phi,\chi)\right)+3H\dot\chi^2-\dot\phi\dP {V_{dm}}\phi=0~.
\end{gather}
Taking the average yields
\begin{gather}
\dot\rho_{dm}+3H\rho_{dm}-\frac 12\dot\phi\dP {M^2(\phi)}\phi\vev{\chi^2}=0~,
\end{gather}
where we have written $\vev{\rho_{dm}}$ as $\rho_{dm}$ and $\vev{\dot\rho_{dm}}$ as $\dot\rho_{dm}$, since the density is not sensible to the oscillations, to a good approximation. This can be easily seen by assuming the rapid oscillations of $\chi(t)$ are described by a sinusoidal function, and hence the density depends only on the amplitude of the oscillations, which is not affected by a cyclic average.

Substituting the average of $\chi^2$ given by Eq. \eqref{avechi}, we obtain
\begin{gather}
\label{avrho}
\dot\rho_{dm}+3H\rho_{dm}=\frac 12\dot\phi\frac{1}{M^2(\phi)}\dP {M^2(\phi)}\phi\rho_{dm}~.
\end{gather}
So, as previously mentioned, the equivalence relation between coupled quintessence and the field theory approach is established via the relationship
\begin{equation}
f(\phi)=\frac 12 \dP {\ln M^2(\phi)}\phi~.
\end{equation}
Furthermore, Eq. \eqref{avrho} can be formally solved as a function of $\phi$, through
\begin{gather}
\rho_{dm}(\phi,a)=n_0a^{-3}M(\phi)~,
\end{gather}
where $n_0$ is an integration constant. Notice this corresponds to the statement that $\rho=nM$, with $M$ being the DM mass and $n$ the number density, proportional to $a^{-3}$. With this solution, the dynamics is reduced to a single differential equation for $\phi$: this can be obtained from Eq. \eqref{phi}, with $V$ being replaced by an effective potential $V_{\text{eff}}$ given by
\begin{gather}
V_{\text{eff}}(\phi,a)=V_{de}(\phi)+\rho_{dm}(\phi,a)~.
\end{gather}

These equations are valid as long as $M^2(\phi)\gg H^2$, otherwise the oscillation regime is not relevant and we must also solve Eq. \eqref{chi}.

\subsection{Modified Potential}

Having defined the potential and derived the relevant equations, we are now ready to draw some general conclusions about the importance of each of the terms of the potential. As will become clear below, such results motivate a modification of the potential.

First, notice that for a sufficiently large $\phi$ we have $M^2(\phi)\approx m^2$. At this regime, examining $\rho_{dm}$ and its derivative, we see that the interaction is irrelevant. On the other hand, for small values of $\phi$, the term with $m^2$ can be neglected. Assuming the polynomial $\tilde P(\phi)$ to be of order one, the transition value $\phi_c$ between the two phases can be estimated by setting $m^2e^{\lambda\phi}=1$, which results in
\begin{equation}
\label{lamb_m}
\phi_c\approx-\frac2\lambda\ln m~.
\end{equation}
Thus, for $\phi<\phi_c$, the interaction is relevant, becoming subdominant as the value of the scalar field grows. We are interested in studying the late time behavior of the Universe, near the stage of accelerated expansion. It is relevant then to estimate whether the interaction is important at late times. The value of the DE field near the present $\phi(0)$ can be estimated by assuming that $\rho_{de0}\approx V_{de}(\phi(0))$ and that it is close to the critical density $\rho_{c0}$, which gives
\begin{gather}
\phi(0)\approx-\frac 1 \lambda \ln \rho_{c0}~.
\end{gather}
Requiring that $\phi_c\geq\phi(0)$, yields a rather low bound for the bare mass,
\begin{gather}
\label{bound}
m\lesssim\sqrt{\rho_{c0}}\sim10^{-60}~.
\end{gather}
Thus, this analysis hints that the effects of the interaction will not be detected at the present unless the DM bare mass is unnaturally small. If $\tilde P(\phi)$ is $\mathcal O (10^s)$, this estimate increases by roughly $2s$ orders of magnitude, which would only shift the naturalness problem to $\tilde P(\phi)$.

Another important situation is the onset of the oscillatory phase: we must establish the $\phi$ field value for which $M^2(\phi)\gtrsim H^2$. In order to obtain it, we use the well-known result \cite{DE_rev} that exponential potentials lead to scaling solutions. Albeit this is not strictly valid in our model, since it is not a pure exponential, we proceed by assuming a scaling behavior before $\phi$ falls in the minimum of the potential, since the polynomial $P(\phi,\chi)$ varies little during that stage. In that case, we have
\begin{gather}
\Omega_{de}\approx \frac{3(w+1)}{\lambda^2}\Rightarrow V_{de} \sim \frac{9(w+1)}{\lambda^2}H^2~,
\end{gather}
in which $w$ is the effective EOS parameter for the combination of all the components. We can then rewrite the oscillation condition as
\begin{gather}
\frac{9(w+1)}{\lambda^2}\frac{m^2 e^{\lambda\phi}+2\tilde P }{P_{\phi} }\gtrsim 1~.
\end{gather}
We recall that if the scaling occurs during nucleosynthesis, then $\lambda\gtrsim10$ \cite{BeaHanMel}. With such a value for $\lambda$ and assuming that $\tilde P\sim P_\phi$, the l.h.s. is always less than unity when the interaction is relevant, meaning that during that stage the field $\chi$ has not begun oscillating; conversely, one expects oscillations to start as the interaction becomes unimportant. In particular, for the threshold mass of Eq. \eqref{bound}, the field may not oscillate until the present, implying the absence of dark matter as such in the Universe in the past. 

Confronted with these problems, we modify the model so to allow for a difference in the behavior of both exponentials, i.e. we choose instead,
\begin{align}
\label{Vlambdabar}
V(\phi,\chi)=&e^{-\lambda\phi}\left(A+(\phi-\phi_0)^2\right)\\
+&e^{-\bar\lambda\phi}\tilde P(\phi)\chi^2+\frac 12 m^2\chi^2\nonumber~.
\end{align}
with $\bar\lambda\neq\lambda$. This relaxes the constraint on $m$ to the less strict condition
\begin{gather}
m\lesssim\rho_{c_0}^{\bar\lambda/2\lambda}~.
\end{gather}
This modification evades the problem associated to the onset of oscillations, as these start while the interaction is still relevant.

A different modification could have been made, by discarding the assumption that the parameters $B$, $C$ and $D$ in $\tilde P(\phi)$ are $\mathcal O (1)$ in terms of $M_p$. However, a solution to the problems mentioned above would require that they are increased by several orders of magnitude: this is rather unnatural, since they are already at the Planck scale.

\section{Numerical Results}\label{Res}

Let us start by rewriting the equations in terms of the number of e-folds $N=\ln a$ and, for convenience, use the rescaled variables of Ref. \cite{Gardner},
\begin{gather}
\tilde H=\frac H{H_0}e^{2N},\ \ \tilde \Phi=\frac {\dot\phi}{H_0}e^{2N},\ \ \tilde X=\frac {\dot\chi}{H_0}e^{2N}~,
\end{gather}
where $H_0=72~\text{km/s/Mpc}$ is the present value of the Hubble constant. Thus, Eqs. \eqref{phi}, \eqref{chi} and \eqref{fried} now read
\begin{align}
&\tilde H^2=\Omega_{m0}e^N+\Omega_{r0}+\frac 16 \tilde\Phi^2+\frac 16 \tilde X^2+\frac{e^{4N}}{3H_0^2}V(\phi,\chi)~,\nonumber \\
&\tilde H(\tilde\Phi'+\tilde\Phi)+\frac{e^{4N}}{H_0^2}\dP {V}\phi=0~,\\
&\tilde H(\tilde X'+\tilde X)+\frac{e^{4N}}{H_0^2}\dP {V}\chi=0\nonumber ~,
\end{align}
where the primes denote derivatives with respect to $N$. These changes improve the numerical robustness of the system by shortening the range of values taken by the new variables. 
From the onset of the oscillatory phase we shall use the averaged equations instead, which become
\begin{align}
&\tilde H^2=\Omega_{m0}e^N+\Omega_{r0}+\frac 16 \tilde\Phi^2+\frac{e^{4N}}{3H_0^2}V_{\text{eff}}(\phi)~,\label{aveeqmov}\\
&\tilde H(\tilde\Phi'+\tilde\Phi)+\frac{e^{4N}}{H_0^2}\dP {V_{\text{eff}}}\phi=0\nonumber~.
\end{align}

We now integrate the equations from $N=-70$ to $N=5$, ranging from the Planck epoch to some time in the future. We present our results in terms of $\log a$ instead of $N$, since it is easier to translate the former into redshift $z$. Nucleosynthesis takes place around $\log a=-10$.

We begin by studying variations of the mass $m$ and later of the exponent $\bar\lambda$. The parameters of the DE potential are kept fixed at $A=0.01$, $\phi_0=28.6$ and $\lambda=9.5$. These are chosen so that there is a local minimum of the potential at the present epoch, while also complying with the bound for $\lambda$ from nucleosynthesis. As long as they verify these conditions, they can be changed without influencing the qualitative results. The parameters of the polynomial $\tilde P(\phi)$ are kept at $B=C=D=1$ in all the plots presented, since no relevant change occurs when they are varied; in fact, even an increase by several orders of magnitude is similar to a change in $m$ or $\bar\lambda$.

\begin{figure}
	\centering
			\includegraphics[width=0.4\textwidth]{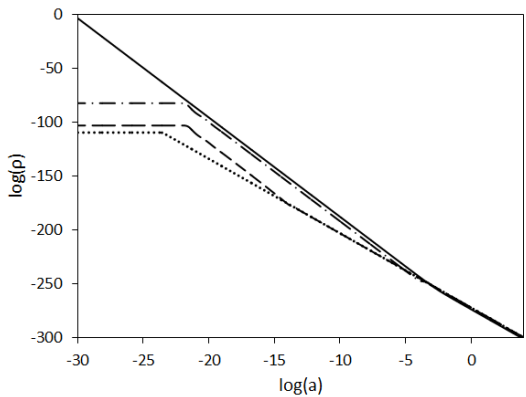}
	\caption{Evolution of $\log\rho_{dm}$ for $m=10^{-55}$ (dot-dashed), $m=10^{-35}$ (dashed) and $m=10^{-15}$ (dotted), which is equivalent to the noninteracting case, as compared to the background density, $\log(\rho_m+\rho_r)$ (solid).}
	\label{rho11}
\end{figure}

\begin{figure}
	\centering
		\includegraphics[width=0.4\textwidth]{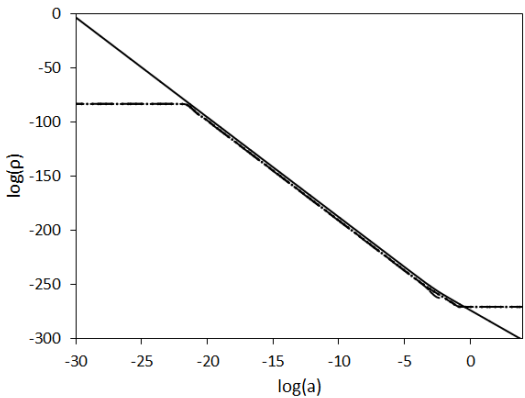}
	\caption{Evolution of $\log\rho_{de}$ for $m=10^{-55}$ (dot-dashed), $m=10^{-35}$ (dashed) and $m=10^{-15}$ (dotted), which is equivalent to the noninteracting scenario, as compared to background density, $\log(\rho_m+\rho_r)$ (solid).}
	\label{rho12}
\end{figure}

We show in Figs. \ref{rho11} and \ref{rho12} the effect of the interaction on $\rho_{dm}$ and $\rho_{de}$ for the indicated values of the bare mass $m$. Notice that the result for $m=10^{-15}$ is equivalent to the noninteracting scenario and for this reason, it is shown for comparison in all figures. This is the range for which we can find initial conditions leading to suitable parameters for the Universe at present and after passing through a DM-dominated phase. We see that in these cases, $\rho_{de}$ is not significantly changed. Moreover, as mentioned above, there is no effect close to the present for the studied range of masses. Nevertheless, the solution for the lowest mass ($m=10^{-55}$) is interesting in that scaling occurs for both DE and DM. In that case, the oscillations do not start until the matter-dominated era, which may be problematic for structure formation. 

\begin{figure}
	\centering
		\includegraphics[width=0.4\textwidth]{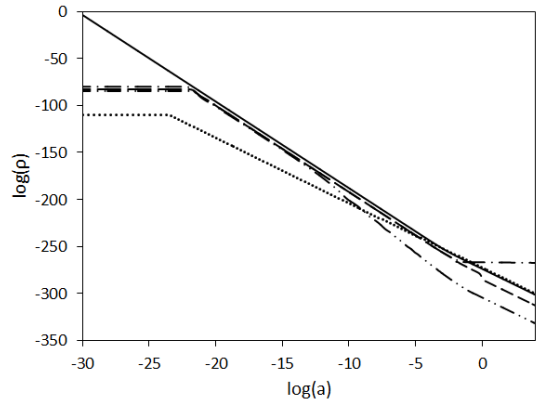}
	\caption{Evolution of $\log\rho_{dm}$ for $m=10^{-60}$ with the initial conditions $\chi_i=1$ (double-dot-dashed), $\chi_i=2.609$ (dashed) and $\chi_i=10$ (dot-dashed), as compared to background density $\log(\rho_m+\rho_r)$ (solid) and to the noninteracting case (dotted).}
	\label{rho21}
\end{figure}

\begin{figure}
	\centering
		\includegraphics[width=0.4\textwidth]{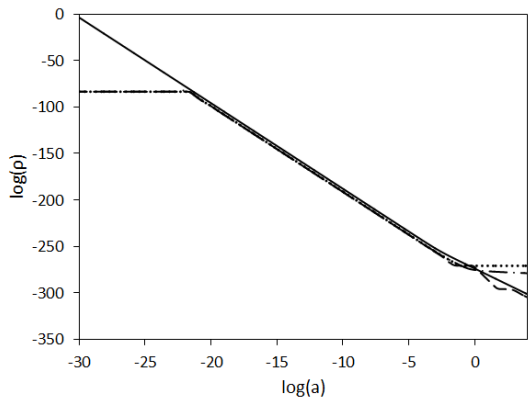}
	\caption{Evolution of $\log\rho_{de}$ for $m=10^{-60}$ with the initial conditions $\chi_i=1$ (dotted), $\chi_i=2.609$ (dashed) and $\chi_i=10$ (dot-dashed), as compared to background density $\log(\rho_m+\rho_r)$ (solid) and to the noninteracting case (superimposed with the $\chi_i=1$, since it is indistinguishable).}
	\label{rho22}
\end{figure}

For much smaller masses, such as the one predicted by Eq. \eqref{bound} ($m=10^{-60}$), this is even more dramatic, as shown in Figs. \ref{rho21} and \ref{rho22}. Two classes of solutions exist with an accelerated expansion at present, both of them unrealistic: in the first case dark matter never dominates (dotted case, Fig. \ref{rho21}), while in the second it dominates, but does not oscillate around the minimum of its potential (dot-dashed case, Fig. \ref{rho21}). In this second type of solution, oscillations never start and the $\chi$ field instead slowly rolls down its potential, thus giving rise to an accelerated expansion. Furthermore, there is a special solution of the first type (dashed) that possesses transient accelerated behavior, although with the same problems already mentioned.

\begin{figure}
	\centering
		\includegraphics[width=0.4\textwidth]{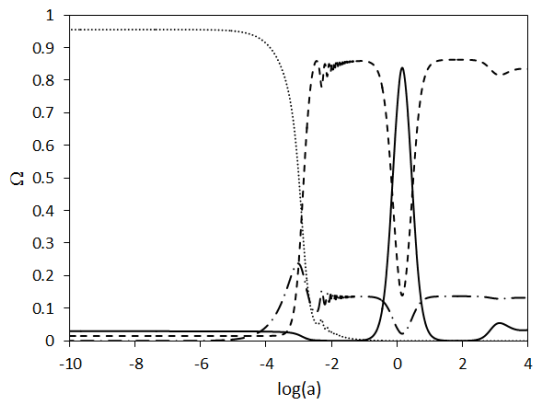}
	\caption{Results for $m=5.9\times 10^{-57}$ showing the evolution of the relative densities $\Omega_{de}$ (solid), $\Omega_{dm}$ (dashed), $\Omega_{r}$ (dotted) and $\Omega_{m}$ (dot-dashed).}
	\label{Om951}
\end{figure}

\begin{figure}
	\centering
		\includegraphics[width=0.4\textwidth]{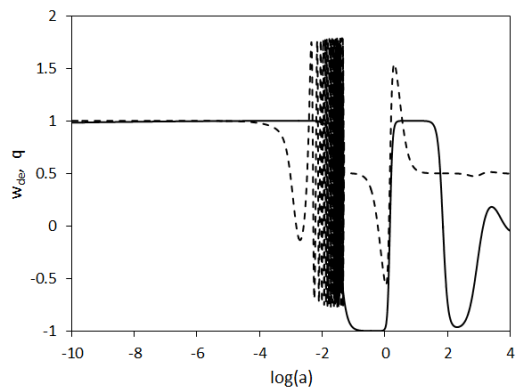}
	\caption{Results for $m=5.9\times 10^{-57}$ showing the evolution of the deceleration parameter $q$ (dashed) and the DE EOS parameter $w_{de}$ (solid). Also shown is the effect of the oscillations on the deceleration parameter before $\log a=-1.3$; at that moment the oscillations are averaged and henceforth the evolution of the relevant quantities is obtained in terms of Eq. \eqref{aveeqmov}.}
	\label{Om952}
\end{figure}

The two regimes, $m\geq10^{-55}$ and $m=10^{-60}$, have a frontier, at $m=5.9\times 10^{-57}$, for which the solution still agrees with the observational constraints and presents a period of transient accelerated expansion, as shown in Figs. \ref{Om951} and \ref{Om952}. Besides being transient, this solution presents other features: it clearly shows the onset of oscillations to be very close to the present--which, in principle, impairs structure formation; $\Omega_{dm}$ is non-negligible during nucleosynthesis, as is $\Omega_{de}$, but their sum is small enough to comply with the bounds in Ref. \cite{BeaHanMel}, $\Omega_r\gtrsim0.95$. Also visible in Fig. \ref{Om952} is the averaging of the oscillations around $\log a=-1.3$. 

\begin{figure}
	\centering
		\includegraphics[width=0.4\textwidth]{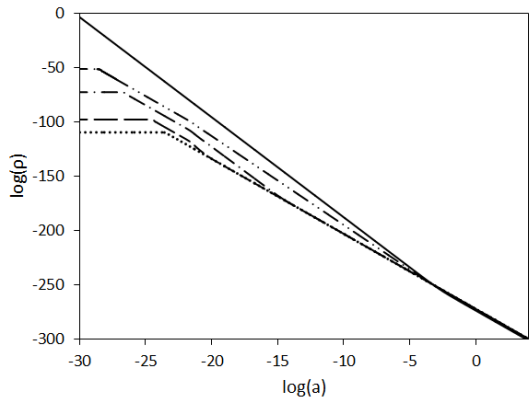}
	\caption{Evolution of $\log\rho_{dm}$ for $m=10^{-15}$ for $\bar\lambda=9.5$ (non-interacting case, dotted), $\bar\lambda=6.5$ (dashed), $\bar\lambda=4.5$ (dot-dashed) and $\bar\lambda=2.8$ (double-dot-dashed), as compared to background density $\log(\rho_m+\rho_r)$ (solid).} 
	\label{rho31}
\end{figure}

\begin{figure}
	\centering
		\includegraphics[width=0.4\textwidth]{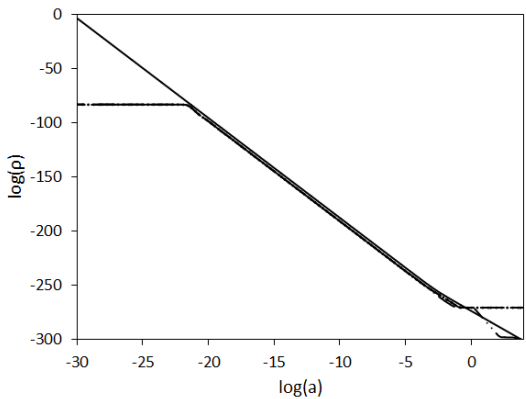}
	\caption{Evolution of $\log\rho_{de}$ for $m=10^{-15}$ for $\bar\lambda=9.5$ (non-interacting case, dotted), $\bar\lambda=6.5$ (dashed), $\bar\lambda=4.5$ (dot-dashed) and $\bar\lambda=2.8$ (double-dot-dashed), as compared to background density $\log(\rho_m+\rho_r)$ (solid).} 
	\label{rho32}
\end{figure}

\begin{figure}
	\centering
		\includegraphics[width=0.4\textwidth]{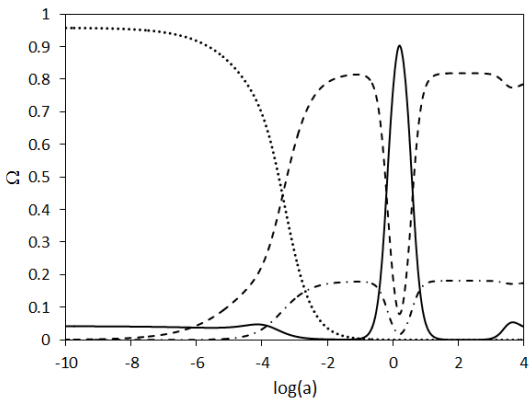}
	\caption{Results for $m=10^{-15}$ and $\bar\lambda=2.8$ showing the evolution of the relative densities $\Omega_{de}$ (solid), $\Omega_{dm}$ (dashed), $\Omega_{r}$ (dotted) and $\Omega_{m}$ (dot-dashed).}
	\label{Om11}
\end{figure}

\begin{figure}
	\centering
		\includegraphics[width=0.4\textwidth]{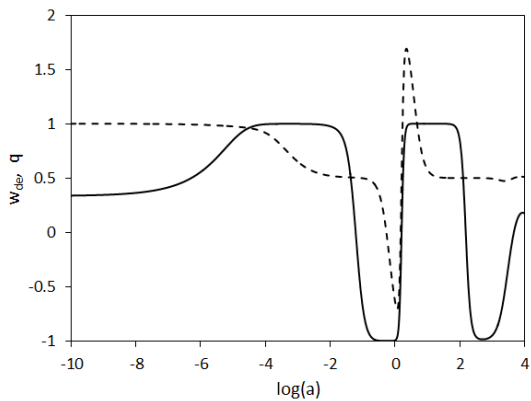}
	\caption{Results for $m=10^{-15}$ and $\bar\lambda=2.8$ showing the evolution of the deceleration parameter $q$ (dashed) and the DE EOS parameter $w_{de}$ (solid).}
	\label{Om12}
\end{figure}

We now turn to the case $\bar\lambda\neq\lambda$, which we study by fixing the mass at $m=10^{-15}\sim 1~\tev$. We obtain somewhat similar results to those already found for masses $m>10^{-55}$ for a large range of $\bar\lambda$, as seen in Figs. \ref{rho31} and \ref{rho32}. As before, there are no relevant effects at the present epoch, with the DM density changing only at early times. However, there is a special case for $\bar\lambda=2.8$, where a solution with transient acceleration is found again, as shown in Figs. \ref{Om11} and \ref{Om12}. This transient solution does not present the problem found previously in the other solution of this kind nor does it have an unnaturally small mass. The only slight anomaly is that $\Omega_{dm}$ starts increasing around $\log(a)=-8$, an effect of the interaction.

\begin{figure}
	\centering
		\includegraphics[width=0.4\textwidth]{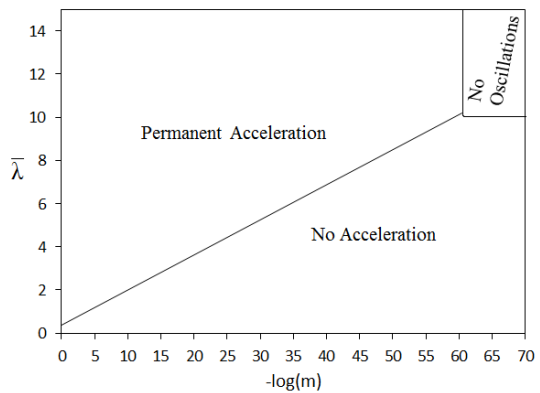}
	\caption{Plot of the parameter space singling out the line where transient acceleration solutions occur and the line where $M^2(\phi_0)=H_0^2$, the limit for oscillations.}
	\label{plot}
\end{figure}

The same transient result is also found for other pairs of $(\bar\lambda,m)$ and their values are plotted in Fig. \ref{plot}. The expression found for the ``transient line'' is
\begin{gather}
\bar\lambda=-0.1625\log(m)+0.3706~.
\end{gather}
This expression is rather similar to Eq. \eqref{lamb_m}, changing $\lambda$ to $\bar\lambda$, since the slope is $-2\ln 10/\phi(0)\approx-2\ln 10/\phi_0=-0.16102$. This is expected, since in these transient solutions we anticipate the interaction to be relevant just until the present. As a consequence, it is not surprising that the smaller the mass, the greater is the required value for $\bar\lambda$.

However, not all of those $(\bar\lambda,m)$ pairs are equally interesting, since in situations with mass smaller than $10^{-15}$, such as $m=10^{-20}$, the DM density parameter starts growing sooner and can exceed the mentioned limit from nucleosynthesis. This trend is not verified for the lowest of masses, as is attested by the solution for $m=5.9\times10^{-57}$.

The explanation for this transient behavior lies in the interaction: the addition of $\rho_{dm}$ to the effective potential raises the minimum of $V_{de}$, thus allowing for the field to escape and continue to roll down the exponential. However, for parameter values below the ``transient line'' of Fig. \ref{plot}, we obtain unrealistic scenarios; for those values the interaction is stronger, so either the solution presents a very low DM density at present or the accelerated phase is nonexistent, both of which are in conflict with observational data.

Albeit appealing, this transient scenario requires some degree of fine-tuning, since the precision required is $\sim 10^{-3}$ in $\bar\lambda$, using bounds on the values of $\Omega_{i0}$ in compliance with \cite{WMAP}. The reason for that is that we need to ``dissolve'' the minimum for the field to run, but still allow for $\phi$ to slow down enough to create an accelerated expansion phase. Such solutions have been found previously in the absence of interaction, for example, in the model of Ref. \cite{AlbSko}, for a similar degree of fine-tuning in $V_{de}$. The present result shows that our model provides an alternative path to achieve such solutions, similarly to the two-field quintessence approach of Ref. \cite{BenBerSan}. The advantage of the present proposal is that it requires tuning only of $\bar\lambda$ or $m$, the initial condition of $\chi$ fixed to reproduce the present dark matter density.

\section{Discussion and Conclusions}\label{Con}

In this paper we have studied a model with two coupled scalar fields, in which one plays the role of dark matter and the other of dark energy at late time. We find solutions that reproduce the current observational data and show that the interaction is irrelevant at the present epoch in the case $\lambda=\bar\lambda$ [cf. Eqs. \eqref{model} and \eqref{Vlambdabar}] for masses larger than $m=5.9\times 10^{-57}\sim10^{-29}\ev$. For this last case the interaction gives rise to a transient stage of acceleration and may be detectable, specially by its effects on structure formation (see e.g. Refs. \cite{OB_FGP_MLD,OB_FGP_MLD2,OB_FGP_MLD3}), an issue that should be addressed in a future work.

The study of the case $\lambda\neq\bar\lambda$ led us to find similar behavior for most of the values of $\bar\lambda$, with an irrelevant contribution of the interaction at late time. The difference appears only for specific values in a line of the parameter space of $\bar\lambda$ and $m$. Those solutions lead to a transient stage of acceleration, not found in the other cases. This stage is compatible with observations and provides a way out to the accelerating regime--a useful property for defining suitable asymptotic states free from future horizons in fundamental theories such as string theory (see Ref. \cite{BenBerSan} and references therein). The ``transient line'' is also the dividing line between solutions with and without an accelerated expansion.

\vskip 2cm
\bibliography{DM_DE_int8}

\end{document}